# Er-doped oxide nanoparticles in silica-based optical fibers

Wilfried Blanc, Bernard Dussardier
*Université de Nice-Sophia Antipolis, LPMC CNRS UMR6622, Parc Valrose, 06108 Nice Cedex 2, France*

Mukul C. Paul
*Fibre Optics Laboratory, Central Glass and Ceramic Research Institute (CGCRI), 196, Raja S.C. Mullick Road, Jadavpur, Kolkata-32, India*

Erbium-doped materials are of great interest in optical telecommunications due to the $Er^{3+}$ intra-4f emission at 1.54 µm. Erbium-Doped Fiber Amplifiers (EDFA) were developed in silica glass because of the low losses at this wavelength and the reliability of this glass. Developments of new rare-earth doped fiber amplifiers aim to control their spectroscopic properties: shape and width of the gain curve, optical quantum efficiency, .... Standard silica glass modifiers, such as aluminum, give very good properties to available EDFA. However, for more drastic spectroscopic changes, more important modifications of the rare-earth ions local environment are required. To this aim, we present a fiber fabrication route creating rare-earth doped calco-silicate or calco-phospho-silicate nanoparticles embedded in silica glass. By adding alkaline earth elements such as calcium, in low concentration, one can obtain a glass with an immiscibility gap. Then, phase separation occurs under an appropriate heat treatment. We investigated the role of two elements: calcium and phosphorus (a standard silica modifier). Under Scanning Electron Microscopy analyses, nanoparticles are observed only when calcium is incorporated. The size of the particles is determined to be around 50 nm in preform samples. The nature of these particles depends on phosphorus content: without P, electron diffraction pattern reveals an amorphous particle while it is partially crystallized when phosphorus is added. Besides, through Energy Dispersive X-Ray techniques, we succeeded to show that erbium ions are located into the nanoparticles.

**Introduction**

Among the rare-earths, erbium is especially interesting because the $^{13}I_{3/2} \rightarrow {}^{15}I_{3/2}$ transition at ≈1.54 µm coincides with the lowest attenuation window of silica glass fiber; hence, erbium doped glasses can be used to amplify attenuated signals in optical fiber telecommunication systems. While Erbium-Doped Fiber Amplifier (EDFA) was developed 20 years ago, extensive studies are still carried out to improve its properties. In particular, linear dimensions should be reduced although the severe limitations imposed by the poor rare earth solubility into silica matrix.. Moreover, spectral bandwidth should be increased to improve the Wavelength Division Multiplexing (WDM) applications. To solve these problems, various approaches based on nanoparticles incorporation have been proposed.

Over the past several decades considerable work have been carried out for incorporation of rare-earth oxide nano-crystallites into different glass hosts. Different processes have been developed such as cosputtering technique[1], pyrolysis[2], ion-implantation[3], laser ablation process[4] and sol-gel process[5]. Another process which was recently developed by Liekki, Finnish company, is the direct nano-particles deposition process[6]. All these processes were related to the outer vapour deposition except the sol-gel process which involves longer time for fabrication of preform. A lot of work was also studied for development of suitable silica and non-silica based glass host for doping of rare-earth ions to improve their lasing and amplification properties . In most of the cases core glass is made of non-silica based or very low silica-based materials made by double crucible furnace melting procedure but they are not very much reliable for the system comparable to the high silica based rare-earth doped fibers. Otherwise, there are reports on thin films containing semiconductor and metal nanoparticles acting as sensitizers, absorbing the incident light with a high cross-section and exciting erbium co-doped ions in the nanoparticle vicinity through energy transfer.

Here we propose for the first time to incorporate Er into oxide nanoparticles in optical fibers to improve the spectroscopic properties of erbium ions. To our knowledge only few studies on nanostrutured silica fibers were dedicated to metal ions properties[7]. In our samples, oxide nanostructures are prepared by adjusting the composition of usual modifiers in the core of the fiber. A glass with an immiscibility gap is obtained and two phases are then formed with high and low silica content. The role of calcium on the preparation of the nanostructures is discussed in this contribution.

## Experimental details

Preforms were fabricated by the conventional Modified Chemical Vapor Deposition (MVCD) technique. In this process, gaseous chlorides ($SiCl_4$, $GeCl_4$, $POCl_3$) are passed through a rotating silica tube, heated by an external burner in translation along the tube. Due to the high temperature, chlorides oxidize, forming particles which deposit on the inner wall. This porous layer turns into a glassy layer when the burner passes over it (the temperature is around 1500 °C). In the final stage, the tube is collapsed into a rod at a temperature higher than 1800°C. This rod can be drawn into fibers. Opto-geometric properties of the perform (or fiber) are determined by adjusting the composition and number of layers. In our samples, phosphorous and germanium concentrations are ~1 mol% and 2 mol%, respectively.

Erbium and calcium ions were incorporated through the solution doping technique. An alcoholic solution (of desired strength of $ErCl_3.6H_2O$ or $CaCl_2.6H_2O$) is soaked for 2 hours in the unsintered core layer. After removing the solution, the layer is dried and sintered. Erbium concentration is estimated through absorption spectra to be around 1000 ppm. Three calcium concentrations in the solution were prepared: 0, 0.01 and 0.1 mol/l. The core diameter was measured with a preform analyzer (York Technology P 101 ) to be 1 mm.

The field emission scanning electron microscope (FESEM) image of the core glass of highly polished preform section of thickness around 2mm was taken. The morphology of such type of core glass was also studied using TEM image of some of the preform samples. The preform TEM specimens were mechanically polished and dimpled to thickness of about 10 μm. The final thinning of the sample to electron transparency was carried out using an Ar ion mill. The TEM examinations were performed with a 200 kV JEM-2000FX transmission electron microscope equipped with Evex ultra-thin window X-ray detector.

## Results and discussion

When the calcium concentration is increased, the color of the central core of the preform turns from transparent to milky. This variation is explained by the structural changes of the core. For preforms with calcium concentration higher than 0.01 mol/l, nanoparticles are observed. A TEM picture of such a preform is presented on figure 1. We can clearly observed polydisperse spherical nanoparticles with a mean diameter of 50 nm. Smaller particles of 10 nm are visible. The size of the biggest particles was around 200 nm (not shown in figure 1). When the calcium concentration decreases, the size distribution of the particles is nearly identical but the density is lowered.

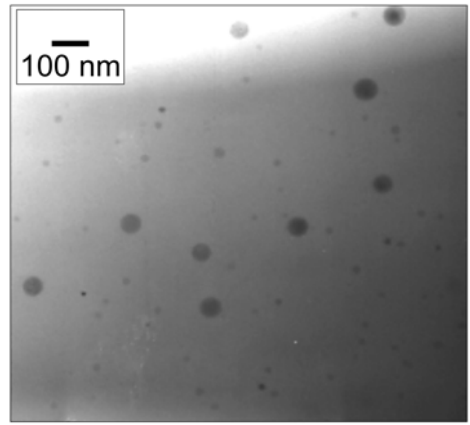

Figure 1: TEM image from sample doped with Ca and P

The composition of the sample was investigated by EDX analyses. When nanoparticles are analyzed, a same amount of Ca, P and Si is found while only Si is detected outside of the particles. Germanium seems to be homogeneously distributed over the entire glass. The structure of the particles was determined with electron diffraction pattern along the TEM analyses. When calcium and phosphorus are incorporated, the oxide particles are amorphous while they are partially crystalline when there is no phosphorus. As the resistance of the structures against the electron diffraction beam irradiation damage is low, the exact nature of the crystal phase could not be determined. When erbium is added to the composition, it is found to be inserted into the particles as it is presented on figure 2.

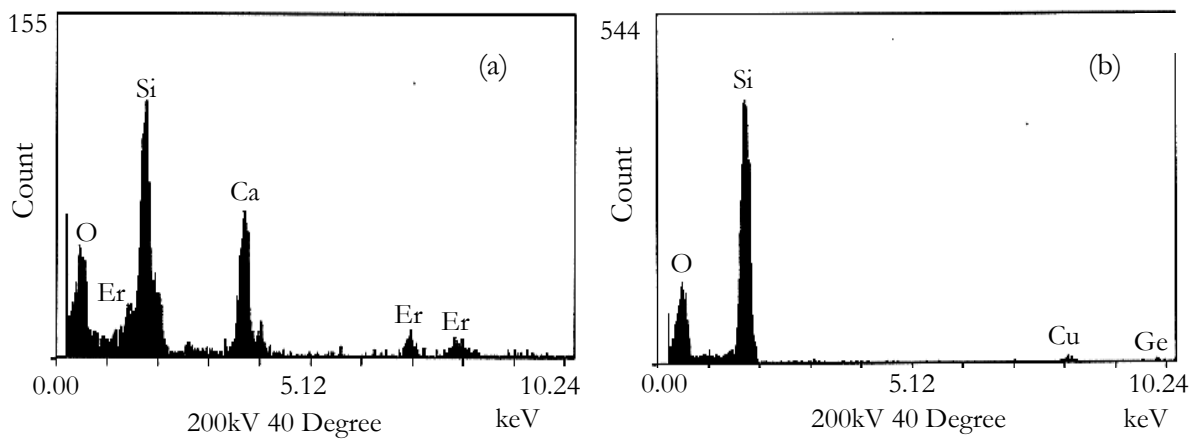

Figure 2: Energy Dispersive X-ray Analyses spectra of the preform sample doped with Ca and Er. The area analyzed corresponds to the nanoparticle (a) and outside (b)

The formation of these nanoparticles can be discussed on the basic principle of the phase separation phenomena and crystal growth mechanisms which normally occur in bulk silica-based glass. On the basis of thermodynamical data such as activity coefficient, entropy of mixing, enthalpy of mixing and Gibbs-free energy change, the phase diagram of the $SiO_2$-CaO binary compound was derived using Factstage software (figure 3). A miscibility gap is found when the CaO concentration is between 2 and 30 mol%. It coincides with the experimentally determined miscibility gap region of binary $SiO_2$-CaO glass[8].

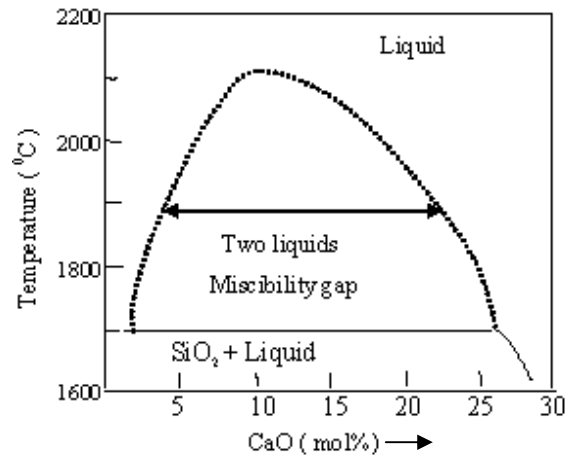

Figure 3: Miscibility-gap in the derived phase-diagram of binary $SiO_2$-CaO glass

The prerequisite for phase separation in multi-component glasses is the existence of an immiscibility region in their phase diagram. Glasses with compositions within the immiscibility region however form a clear and transparent glass when their melts are quenched rapidly below the glass transition temperature, $T_g$. Below this temperature the viscosity is too high to allow for the necessary rearrangements of structural groups required for phase separation. However when such glasses are reheated to temperatures in the range between $T_g$ and the upper critical temperature of the immiscibility dome, sub-liquidus phase separation occurs. Two principal mechanisms of phase separation have been proposed so far, phase separation by spinodal decomposition and by nucleation and growth. Due to the spherical shape of the particles, one may think that the last mechanism prevails[9]. The opaqueness observed when calcium concentration increases is a consequence of light scattering. This induces high loss and has to be avoided for telecommunication applications. Tick discussed parameters to be controlled and mentioned that the particle size must be less than 15 nm and the particle-size distribution must be narrow[10].

**Conclusion**

In this contribution we have investigated a new route to prepare erbium-doped oxyde nanoparticles into preforms. By adding calcium into silica-based glass, we successfully prepare nanostructures through phase separation. The mean diameter is around 50 nm with a high calcium and phosphorus content. Erbium ions are found to be located into nanoparticles. This would lead to improve spectroscopic properties of the rare-earth and benefit to the telecommunication systems.

**Acknowledgements**

This work was realized in the frame of a P2R project (Programme de Recherche en Réseau, CNRS (France) and DST (India)). We thank Ph. Vennéguès at CRHEA (CNRS, Valbonne, France) and S.C. Cheng (Corning, New-York, USA) for the TEM images and EDX analyses.